\documentclass[12pt]{iopart}

\usepackage{graphicx}
\usepackage{amssymb}
\usepackage{iopams}
\usepackage{epsfig}

\begin{document}

\newtheorem{theo}{Theorem}

\title{Thermodynamics with generalized ensembles: The class of dual orthodes}
\author{Michele Campisi}
%\email{campisi@unt.edu}
\ead{campisi@unt.edu}
%\affiliation{Department of Physics,University of North Texas, P.O. Box 311427, Denton, TX 76203-1427,USA}
\address{Department of Physics,University of North Texas, P.O. Box
311427, Denton, TX 76203-1427,USA}
\date{\today}

\begin{abstract}
We address the problem of the foundation of generalized ensembles
in statistical physics. The approach is based on Boltzmann's
concept of orthodes. These are the statistical ensembles that
satisfy the heat theorem, according to which the heat exchanged
divided by the temperature is an exact differential. This approach
can be seen as a mechanical approach alternative to the well
established information-theoretic one based on the maximization of
generalized information entropy. Our starting point are the
Tsallis ensembles which have been previously proved to be
orthodes, and have been proved to interpolate between canonical
and microcanonical ensembles. Here we shall see that the Tsallis
ensembles belong to a wider class of orthodes that include the
most diverse types of ensembles. All such ensembles admit both a
microcanonical-like parametrization (via the energy), and a
canonical-like one (via the parameter $\beta$). For this reason we
name them ``dual''. One central result used to build the theory is
a generalized equipartition theorem. The theory is illustrated
with a few examples and the equivalence of all the dual orthodes
is discussed.
\end{abstract}

%\pacs{}

 \maketitle

\section{\label{sec:intro}Introduction}
Around 20 years ago a generalization of the standard
Gibbs-Boltzmann statistics was proposed by C. Tsallis
\cite{Tsallis88}. The generalization scheme proposed by Tsallis
was based on the adoption of a generalized Shannon's informational
entropy, which, when maximized under the constraints of
normalization and average energy, leads to power-law statistics.
Since then, the corresponding generalized thermostatistics, also
named non-extensive thermodynamics, has proved to be a very
powerful tool of investigation within the most diverse fields of
physics.

In two previous works \cite{Campisi07,Campisi06.2} we have
proposed an alternative \emph{mechanical} approach to study the
foundation of the so-called Tsallis ensemble. Those studies
revealed two interesting facts. First, Tsallis ensembles are
\emph{exact orthodes} \cite{Campisi07}. According to Boltzmann's
original definition (see \cite{Gallavotti} for a modern
exposition) these are the ensembles that satisfy the ``heat
theorem''
\begin{equation}\label{}
    \frac{\delta Q}{T} = exact \quad differential
\end{equation}
where $\delta Q = dU +PdV$ and $T$ are calculated from the
ensemble average of properly chosen functions of the microscopic
state (i.e. the 6N-dimensional phase space vector). In this sense
it is said that the orthodes provide good mechanical models of
thermodynamics \cite{Gallavotti}. Second, not only the canonical
ensemble is a special case of Tsallis ensemble, but the
microcanonical is also. That is, the family of Tsallis ensembles
parameterized by the extensivity index $q$ interpolates between
canonical and microcanonical ensemble \cite{Campisi06.2}.

We now have a continuous family of ensembles, namely the Tsallis
ensembles, parameterized by the extensivity index $q$ that
generalizes both the canonical and microcanonical ensembles. All
these ensembles are orthodes. Here we shall generalize further and
show that said family is a subset of a more general class of
orthodes which we shall call \emph{dual orthodes}. The main
ingredient that will be used to develop the generalization is the
\emph{duality} property which characterizes the Tsallis ensemble.
By duality we mean that the ensemble can be parameterized either
through the average energy (microcanonical-like parametrization)
or through the parameter $\beta$ (canonical-like parametrization).
In section \ref{sec:rev} we provide the definition of orthode and
we review the content of Refs. \cite{Campisi07,Campisi06.2}. We
move to propose the generalization in Sec. \ref{sec:general} where
the concept of \emph{dual statistics} is introduced and a
\emph{generalized equipartition theorem} is presented. Section
\ref{sec:idealGas} shows that the standard ideal gas
thermodynamics follows from any dual statistics. In section
\ref{sec:examples} we illustrate the concept of dual ensembles by
studying a few particular cases, which include gaussian and
Fermi-like ensembles. Section \ref{sec:conclusions} gives the
concluding remarks.
%\newpage

\section{\label{sec:rev}The Tsallis orthodes}
This section is devoted to review the results of Refs.
\cite{Campisi07,Campisi06.2}. These concern the foundation of
Tsallis statistics from the viewpoint of \emph{orthodicity}. The
approach can be considered a \emph{mechanical} approach, as
opposed to the standard \emph{information theoretic} one used by
Tsallis \cite{Tsallis88}. The concept of orthode was first
proposed by Boltzmann, but did not meet the same success of other
ideas developed by him, such as the counting method or the
H-theorem. Nonetheless the method provides of a very
straightforward way to asses whether a given statistical ensemble
provides a mechanical model of thermodynamics. This method was
used by Boltzmann to give a theoretical basis for the canonical
and microcanonical ensemble in statistical mechanics. Now we know
that Tsallis ensembles also provide good models of thermodynamics
as well \cite{Campisi07}. The idea on which orthodes are based is
quite simple. Consider a generic statistical ensemble of
distributions, $\rho(\mathbf{z};\lambda_0,\lambda_i)$
parameterized by one ``internal'' parameter $\lambda_0$ and a
given number of ``external'' parameters $\lambda_i$ $i>0$. The
symbol $\mathbf{z}$ denotes the phase space point vector. We
assume the system to be Hamiltonian with $\mathbf{z} = (p_1, p_2,
... ,p_n, q_1, q_2, ... ,q_n)$. The Hamilton function is assumed
to be of the form $H= K(\mathbf{p})+
\varphi(\mathbf{q};\lambda_i)$, $i>0$. The Hamiltonian depends
explicitly on the external parameters, which are also called
``generalized displacements''. For sake of simplicity, we consider
only one external parameter $\lambda_1=V$. This may be, for
example, the volume of a vessel containing the system. The kinetic
energy is assumed to be of the form $K(\mathbf{p})=
\frac{\mathbf{p}^2}{2m}$. Then define the macroscopic state of the
system by the set of following quantities:
\begin{equation}{\label{eq:statedef0}}
\begin{tabular}{l}
$U_\rho \doteq \left\langle H \right\rangle_\rho $\quad``energy'' \\
$T_\rho \doteq \frac{2\left\langle K\right\rangle_\rho}{n}$\quad
``doubled kinetic energy per degree of freedom''\\
$V_\rho \doteq \left\langle V \right\rangle_\rho $ \quad ``generalized displacement'' \\
$P_\rho \doteq \left\langle -\frac{\partial H}{\partial
V}\right\rangle_\rho$ \quad ``generalized conjugated force'',
\end{tabular}  \end{equation}
Now let the parameters $\lambda_0,V$ change of infinitesimal
amounts, and calculate the corresponding change in the macroscopic
state. If the state changes in such a way that the fundamental
equation of thermodynamics, namely
\begin{equation}\label{eq:heatTheo}
    \frac{dU_\rho+P_\rho dV_\rho}{T_\rho} = exact\quad
    differential,
\end{equation}
holds, then we can say that the ensemble provides a good
mechanical model of thermodynamics. Equation (\ref{eq:heatTheo})
is also known as the \emph{heat theorem} and expresses the fact
that the heat differential $\delta Q$ admits an integrating factor
($T^{-1}$), whose physical interpretation is that of reciprocal
temperature (namely doubled average kinetic energy per degree of
freedom). Correspondingly the function which generates such exact
differential is interpreted as the physical entropy of the system.
The ensembles that satisfy Eq. (\ref{eq:heatTheo}) were called
\emph{orthodes} by Boltzmann \cite{Bol84} (see also
\cite{Gallavotti} and \cite{Campisi05} for modern expositions).
Boltzmann proved that microcanonical and canonical ensembles are
\emph{orthodes}, and placed this fact at the very heart of
statistical mechanics. For example, the canonical ensemble is
parameterized by one internal parameter, usually indicated by the
greek letter $\beta$, and the external parameter $V$:
\begin{equation}\label{eq:canonical-ensemble}
\rho_c(\mathbf{z};\beta,V)= \frac{e^{-\beta
H(\mathbf{z};V)}}{Z(\beta,V)}
\end{equation}
where
\begin{equation}\label{}
Z(\beta,V) \doteq \int d\textbf{z} e^{-\beta H(\mathbf{z};V)},
\end{equation}
The following function:
\begin{equation}\label{Eq:canonicalEntropy}
    S_c(\beta,V) = \beta U_c(\beta,V) + \log Z (\beta,V)
\end{equation}
generates the heat differential, therefore is the entropy
associated with the canonical orthode. To prove that, we have to
calculate the partial derivatives of $S_c$:
\begin{equation}\label{}
    \frac{\partial S_c}{\partial \beta} = U_c + \beta \frac{\partial U_c}{\partial
    \beta} - \left\langle H\right\rangle_c = \beta \frac{\partial U_c}{\partial
    \beta}
\end{equation}
\begin{equation}\label{}
    \frac{\partial S_c}{\partial V} = \beta \frac{\partial U_c}{\partial
    V} - \beta \left\langle \frac{\partial H}{\partial V}\right\rangle_c
    = \beta \frac{\partial U_c}{\partial V} + \beta P_c
\end{equation}
where the symbol $<\cdot>_c$ denotes average over the canonical
distribution (\ref{eq:canonical-ensemble}) and the state
definition (\ref{eq:statedef0}) has been used. Further according
to the canonical equipartition theorem
\begin{equation}\label{eq:equiTeo-canonical}
   \left\langle p_i\frac{\partial H}{\partial p_i} \right\rangle_c = \frac{1}{\beta}
\end{equation}
Therefore, by comparison with (\ref{eq:statedef0}) $T_c =
\frac{1}{\beta}$. Combining all together we get:
\begin{equation}\label{}
    dS = \beta \left(\frac{\partial U_{c}}{\partial \beta}d\beta + \frac{\partial U_{c}}{\partial V}dV +
    P_c    dV \right)= \frac{dU_c + P_c dV_{c}}{T_c}.
\end{equation}

The microcanonical ensemble is parameterized by $(U,V)$:
\begin{equation}\label{eq:microcanonical-ensemble}
\rho_{mc}(\mathbf{z};U,V)=
\frac{\delta(U-H(\mathbf{z};V))}{\Omega(U,V)}
\end{equation}
where $\delta$ denotes Dirac's delta function and
\begin{equation}\label{eq:Omega}
\Omega(U,V) \doteq \int d\mathbf{z}\delta(U-H(\mathbf{z};V))
\end{equation}
denotes the density of states or structure function
\cite{Khinchin}. As shown in Ref. \cite{Campisi05} the
corresponding entropy is given by
\begin{equation}\label{}
    S_{mc}(U,V) = \log \int d\mathbf{z}\theta(U-H(\mathbf{z};V))
\end{equation}
where $\theta$ denotes Heaviside step function. In this case the
proof is based on the microcanonical equipartition theorem
\cite{Khinchin}:
\begin{equation}\label{eq:equiTeo-microcanonical}
    T_{mc} = \left\langle p_i\frac{\partial H}{\partial p_i} \right\rangle_{mc} = \frac{\Omega(U,V)}{\Phi(U,V)}
\end{equation}
where $<\cdot>_{mc}$ denotes average over the microcanonical
distribution (\ref{eq:microcanonical-ensemble}) and
\begin{equation}\label{}
\Phi(U,V) \doteq \int d\mathbf{z}\theta(U-H(\mathbf{z};V)).
\end{equation}
A proof that the heat theorem is satisfied by the microcanonical
ensemble appeared in \cite{Campisi05}.

The Tsallis ensembles of indices $q \leq 1$ are \emph{dual}
ensembles \cite{Campisi07,Campisi06.2}, namely they can be
parameterized either through $(\beta,V)$ or $(U,V)$:
\begin{equation}\label{eq:Tsallis-ensemble}
    \rho_q (\textbf{z}; \square ,V) = \frac{\left[ 1 -
\beta(1-q)(H(\textbf{z};V)-U)\right]^{\frac{q}{1-q}}}{N_q},
\end{equation}
where
\begin{equation}\label{eq:EN}
    N_q(\square,V) = \int d\textbf{z} \left[ 1 -
(1-q)\beta(H(\textbf{z};V)-U)\right]^{\frac{q}{1-q}}.
\end{equation}
The symbol ``$\square$'' has to be replaced by either $U$ or
$\beta$ according to the parametrization adopted. In the $(U,V)$
parametrization, one first fixes $U$ and then adjusts $\beta$ in
such a way that:
\begin{equation}\label{eq:U=<H>}
   U= \left\langle H \right\rangle_{\rho_q},
\end{equation}
In this parametrization Eq. (\ref{eq:U=<H>}) defines the function
$\beta(U,V)$. In the $(\beta,V)$ parametrization, $\beta$ is fixed
instead and $U$ is adjusted accordingly, in such a way that Eq.
(\ref{eq:U=<H>}) defines the function $U(\beta,V)$. As shown in
Ref. \cite{Campisi07}, no matter the parametrization adopted the
ensembles (\ref{eq:Tsallis-ensemble}) are \emph{orthodes} and the
corresponding entropies are given by:
\begin{equation}\label{eq:S=logEN}
    S^{[q]}(\square,V) = \log
     \int d\textbf{z} \left[ 1 -
(1-q)\beta(H(\textbf{z};V)-U)\right]^{\frac{1}{1-q}}
\end{equation}
The proof of orthodicity is based on the corresponding Tsallis
equipartition theorem: \cite{Martinez02}:
\begin{equation}\label{eq:T=1/beta-tsallis}
    T^{[q]} = \left\langle p_i\frac{\partial H}{\partial p_i} \right\rangle = \frac{1}{\beta}\frac{N_q(\square,V)}{\mathcal{N}_q(\square,V)}
\end{equation}
where:
\begin{equation}\label{}
    \mathcal{N}_q(\square,V) = \int d\textbf{z} \left[ 1 -
(1-q)\beta(H(\textbf{z};V)-U)\right]^{\frac{1}{1-q}}
\end{equation}
The proof of orthodicity of Tsallis ensemble appeared in Ref.
\cite{Campisi07}.

The fact that Tsallis ensembles are orthodes reveals that they are
as well founded as the microcanonical and canonical ensembles, at
least as far as providing good mechanical models of
thermodynamics. The work of Ref. \cite{Campisi06.2} has shown that
the family of Tsallis ensembles parameterized by the index $q$
interpolates between canonical and microcanonical ensembles. In
fact, the former is recovered in the limit $q\rightarrow 1$, and
the latter is recovered in the limit $q \rightarrow -\infty$. This
interpolation is not limited to the distributions only but also to
the corresponding entropies and equipartition theorems. Physically
the Tsallis statistics describes the situation of a system in
contact with a \emph{finite heat bath}, namely, a bath with a
given finite heat capacity \cite{Campisi06.2,Almeida01}:
\begin{equation}\label{eq:CV-q}
    C_V = \frac{1}{1-q}
\end{equation}
As $q$ goes to 1, the heat capacity goes to infinity and
correspondingly one obtains the canonical ensemble (the system is
thermalised). As $q$ goes to $-\infty$, the heat capacity goes to
zero and one recovers the microcanonical situation (the system is
thermally isolated) \cite{Campisi06.2}.

\section{\label{sec:general}Generalization: Dual orthodes}
So far we have seen how the Tsallis ensembles of index $q$ are
orthodes with a special \emph{duality} property. We have also
mentioned that the microcanonical and canonical ensembles are two
special instances of Tsallis ensembles. Now we shall see that the
property of orthodicity can be proved for a general class of
ensembles which share with the Tsallis ensemble the fact that both
$U$ and $\beta$ appear explicitly in their expression. They can be
considered as parameterized by either $U,V$ or $\beta,V$,
depending on which parameter is kept fixed and which one is
adjusted, in such a way that $U=<H>$. We shall call these
ensembles \emph{dual ensembles}.

The generalization is purely formal, in the sense that we shall
assume that all integrals and derivatives written exist. Further
we shall assume that, for given $U,V$ (or $\beta,V$), the equation
$U=<H>$ admits a solution $\beta(U,V)$ (or $U(\beta,V)$) which is
of class $C^1$. These are conditions that must be checked, a
posteriori, on a case by case basis, depending on the explicit
form of the Hamiltonian and of the distribution. Thus, let us
consider a generic ensemble of the form:
\begin{equation}\label{eq:rho=f/N}
\rho(\textbf{z};U,V) =\frac{f[\beta(U-H(\textbf{z};V))]}{G(U,V)},
\end{equation}
Where $G$ and $\beta$ are assumed to be differentiable functions
of $U,V$. This means that we are adopting the $U,V$
parametrization, alternatively we could have adopted the $\beta,V$
parametrization. The values of $G$ and $\beta$ are fixed by the
two constraints
\begin{equation}\label{eq:G=int-f}
    G(U,V) = \int d\textbf{z} f[\beta(U-H(\textbf{z};V))]
\end{equation}
and
\begin{equation}\label{}
    U = \int d\textbf{z} H(\textbf{z};V) \rho(\textbf{z};U,V)
\end{equation}
Throughout this section the symbol $<\cdot>_\rho$ denotes average
over the dual distribution (\ref{eq:rho=f/N}). Consider now a
differentiable function $F(x)$ such that $F(x)>0,F'(x)= f(x)$, and
define the following function:
\begin{equation}\label{eq:S=logMathcalG}
S_\rho(U,V) \doteq \log \mathcal{G}(U,V)
\end{equation}
where
\begin{equation}\label{eq:mathcalG=intF}
\mathcal{G}(U,V) \doteq \int d\textbf{z}
F[\beta(U-H(\textbf{z};V))]
\end{equation}
and all the integrals are extended to the definition domain
$\mathcal{D}$ of the distribution $\rho$. We shall also assume
that a cut-off condition exists such that $F$ is null on the
boundary of the domain:
\begin{equation}\label{eq:F-cutoff}
    F\left[\beta(U-H(\textbf{z};V))\right]_{\textbf{z} \in \partial
    \mathcal{D}}=0.
\end{equation}
Let us define the macroscopic state:
\begin{equation}{\label{eq:statedef}}
\begin{tabular}{l}
$U_\rho \doteq \left\langle H \right\rangle_\rho $ = U \\
$T_\rho \doteq \frac{2\left\langle K\right\rangle_\rho}{n}$ \\
$V_\rho \doteq V $ \\
$P_\rho \doteq \left\langle -\frac{\partial H}{\partial
V}\right\rangle_\rho$
\end{tabular}  \end{equation}

Before proving that the ensembles of the form (\ref{eq:rho=f/N})
are orthodes let us state the following \emph{generalized
equipartition theorem}:
\begin{theo}
In the $U,V$ parametrization the average of $p_i \frac{\partial
H}{\partial p_i}$ is:
\begin{equation}\label{eq:genEquiTheoUV} \left\langle p_i
\frac{\partial H}{\partial p_i}
 \right\rangle_\rho = \frac{1}{\beta(U,V)}
 \frac{\mathcal{G}(U,V)}{G(U,V)}.
\end{equation}
In the $\beta,V$ parametrization the average of $p_i
\frac{\partial H}{\partial p_i}$ is:
\begin{equation}\label{eq:genEquiTheoBETAV} \left\langle p_i
\frac{\partial H}{\partial p_i}
 \right\rangle_\rho = \frac{1}{\beta} \frac{\mathcal{G}(\beta,V)}{G(\beta,V)}
\end{equation}
\end{theo}
The proof is provided in \ref{appSec:gen-equi-theo}. It involves
writing the integral expression of the quantity $\left\langle p_i
\frac{\partial H}{\partial p_i}
 \right\rangle_\rho$, integrating by parts over $p_i$ and using the cut-off condition
(\ref{eq:F-cutoff}). The proof structure is the same as the
structure of the proof of Tsallis equipartition theorem of Ref.
\cite{Martinez02}.

Comparing with the macroscopic state definition
(\ref{eq:statedef}), the generalized equipartition theorem can be
reexpressed as the following compact formula :
\begin{equation}\label{eq:gen-equi-teo}
    T_\rho = \frac{1}{\beta} \frac{\mathcal{G}}{G}
\end{equation}
where it is intended that Eq. (\ref{eq:genEquiTheoUV}) is used in
the $U,V$ parametrization and Eq. (\ref{eq:genEquiTheoBETAV}) in
the $\beta,V$ one.

Equation (\ref{eq:gen-equi-teo}) tells that for generic dual
statistics the quantity $\beta$ \emph{might not} coincide with the
inverse physical temperature. Let us now evaluate the partial
derivatives of the entropy function in Eq.
(\ref{eq:S=logMathcalG}):
 \begin{eqnarray}\label{eq:partialSpartialU}
    \frac{\partial S_\rho}{\partial U_\rho}  &=& \frac{1}{\mathcal{G}}
    \frac{\partial}{\partial{U}}\int
d\textbf{z} F(\beta(U-H))\nonumber \\
    &=& \frac{1}{\mathcal{G}}\int d\textbf{z} f(\beta(U-H)) \left[\frac{\partial
\beta}{\partial U} (U-H) + \beta
    \right]\nonumber \\
    &=& - \frac{G}{\mathcal{G}} \frac{\partial \beta}{\partial U} \langle H-U \rangle_\rho +
    \frac{G}{\mathcal{G}} \beta = \frac{1}{T_\rho}.
\end{eqnarray}
In order to obtain the last equality we used the first definition
in (\ref{eq:statedef}) and Eq. (\ref{eq:gen-equi-teo}).
 \begin{eqnarray}\label{eq:partialSpartialV}
    \frac{\partial S_\rho}{\partial V_\rho} &=& \frac{1}{\mathcal{G}}
    \frac{\partial}{\partial V}\int
d\textbf{z} F(\beta(U-H))\nonumber \\
    &=& \frac{1}{\mathcal{G}}\int d\textbf{z} f(\beta(U-H)) \left[\frac{\partial
\beta}{\partial V} (U-H) - \beta \frac{\partial H}{\partial V} \nonumber \right]\\
    &=& \frac{G}{\mathcal{G}} \frac{\partial \beta}{\partial V} \langle U-H \rangle_\rho
    -
    \frac{G}{\mathcal{G}}\beta   \left\langle \frac{\partial H}{\partial V} \right\rangle_\rho
\nonumber \\
     &=& \frac{P_\rho}{T_\rho}.
\end{eqnarray}
In order to obtain the last equality we used the first and fourth
definitions in (\ref{eq:statedef}) and Eq.
(\ref{eq:gen-equi-teo}). From Eq.s (\ref{eq:partialSpartialU}) and
(\ref{eq:partialSpartialV}) we get:
\begin{equation}\label{}
    dS_\rho= \frac{dE_\rho+P_\rho dV_\rho}{T_\rho}
\end{equation}
Therefore the differential $\frac{dE_\rho+P_\rho dV_\rho}{T_\rho}$
is exact and the entropy is given by Eq. (\ref{eq:S=logMathcalG}).
This implies that the ensembles of the form (\ref{eq:rho=f/N}) are
orthodes, namely they provide good mechanical models of
thermodynamics. In \ref{appSec:orthodicity} we provide a proof
that the heat theorem is satisfied also if the alternative $\beta,
V$ parametrization is adopted. The proof is essentially the same
as for the Tsallis case (see \cite{Campisi07}). Thus, we have
found that the class of orthodes, whose known representatives have
been for more than one century only a few (canonical,
microcanonical, grand-canonical and pressure ensemble
\cite{Gallavotti}) is indeed quite vast and can include other
statistics.

\subsection{\label{subsec:special}Recovery of known cases}
\subsubsection{Canonical}The canonical ensemble is a very special case of dual
orthode where the parameter $U$ does not appear explicitly in the
expression of the distribution. This case is obtained with the
choice:
\begin{equation}
\nonumber  f(x) = F(x) = e^x.\\
\end{equation}
In this case we get
\begin{equation}\label{eq:can-ens}
\rho(\textbf{z};\beta,V) = \frac{e^{\beta(U-H)}} {\int
d\textbf{z}e^{\beta(U-H)}} = \frac{e^{-\beta H}} {\int
d\textbf{z}e^{-\beta H}}.
\end{equation}
The average energy $U$ cancels in the last term of
(\ref{eq:can-ens}). In this sense we refer to the canonical
ensemble as a case of ``hidden dual orthode''. The canonical
entropy is recovered by taking the natural logarithm of
$\mathcal{G}= \int d\textbf{z}e^{\beta(U-H)}$ :
\begin{equation}\label{}
    S(\beta,V) = \beta U + \log \int d\textbf{z} e^{-\beta H}
    \nonumber
\end{equation}
Note also that, from Eq.s (\ref{eq:G=int-f}) and
(\ref{eq:mathcalG=intF}), $G= \mathcal{G}$ in this specific case,
therefore the generalized equipartition theorem
(\ref{eq:gen-equi-teo}) gives $T_\rho = \frac{1}{\beta}$. In this
way the canonical equipartition theorem
(\ref{eq:equiTeo-canonical}) is recovered too.

\subsubsection{Microcanonical}
The microcanonical ensemble is recovered with the choice
\begin{eqnarray}\label{eq:micro-can-ens}
\nonumber    f(x) &=& \delta (x)\\
\nonumber    F(x) &=& \theta (x)
\end{eqnarray}
From the properties of the Dirac delta the distribution in Eq.
(\ref{eq:rho=f/N}) is:
\begin{equation}\label{}
\rho(\textbf{z};U,V) = \frac{\delta(\beta(U-H))} {\int
d\textbf{z}\delta(\beta(U-H))} = \frac{\delta(U-H)} {\int
d\textbf{z}\delta(U-H)}
\end{equation}
As with the canonical case (\ref{eq:can-ens}), the last term in
(\ref{eq:micro-can-ens}), does not depend explicitly on $\beta$,
hence the microcanonical case is also a case of ``hidden dual
statistics''. The microcanonical equipartition theorem
(\ref{eq:equiTeo-microcanonical}) is also recovered. From
(\ref{eq:gen-equi-teo}) one gets:
\begin{equation}\label{}
T_\rho = \frac{1}{\beta}\frac{\mathcal{G}}{G} = \frac{\int
d\textbf{z}\theta(U-H)}{\int d\textbf{z}\delta(U-H)} =
\frac{\Phi}{\Omega}
\end{equation}
where we have used the relations $\theta(a x) = \theta(x)$ (for $a
>0$) and $\delta(a x) = a^{-1}\delta(x)$.

\subsubsection{Tsallis} The Tsallis distribution is recovered with the
choice
\begin{eqnarray}
f(x) &=& \left[1+(1-q)x \right]^{\frac{q}{1-q}} \label{eq:q-exponential}\\
\nonumber  F(x) &=& \left[1+(1-q)x \right]^{\frac{1}{1-q}} \\
\end{eqnarray}
In this case one finds $\mathcal{G} = \mathcal{N}_q$ and $G=N_q$
in Eq.s (\ref{eq:G=int-f}) and (\ref{eq:mathcalG=intF}), so from
Eq. (\ref{eq:gen-equi-teo}) the Tsallis equipartition theorem
(\ref{eq:T=1/beta-tsallis}), is recovered:
\begin{equation}\label{eq:recoveryTsalEquiTeo}
    T_\rho = \frac{1}{\beta}\frac{\mathcal{G}}{G} =
    \frac{1}{\beta}\frac{\mathcal{N}_q}{N_q}.
\end{equation}
Canonical and microcanonical cases are both included in the family
of Tsallis distributions as special cases corresponding to the
values $q=1$ and $q= -\infty$ \cite{Campisi07}.

From \cite{Campisi07,Martinez02} we have for finite $q$'s
$\mathcal{N}_q=N_q$ , therefore Eq. (\ref{eq:recoveryTsalEquiTeo})
becomes:
\begin{equation}\label{eq:equi-teo-tsal}
    T^{[q]} = \frac{1}{\beta} \qquad |q|<\infty
\end{equation}

\section{\label{sec:idealGas}Derivation of the ideal gas thermodynamics}
In the ideal gas case the potential energy $\varphi(\mathbf{z};V)$
is a box potential which constrains the coordinates to lye in an
interval of measure $L=V^{\frac{1}{3}}$ where we assume for
simplicity a cubic box of volume $V$. The box potential reduces
the integration over the configuration space to a domain of
measure $V^{n/3}$, where $n= 3N$ is the total number of degrees of
freedom and $N$ is the number of particles. The Hamiltonian is
purely kinetic $H= \sum_{i=1}^{3N}\frac{p_i^2}{2m}$. Assuming that
the function $\beta(U,V)$ exists, the equation of state is
obtained from Eq. (\ref{eq:partialSpartialV}), namely
$\frac{\partial S_\rho}{\partial V} = \frac{P_\rho}{T_\rho}$:
\begin{eqnarray}\label{}
\frac{\partial S_\rho}{\partial V} &=& \frac{1}{\mathcal{G}}
    \frac{\partial}{\partial V}
   \int_0^{V^{1/3}} d^n\mathbf{q} \int
   d^n\mathbf{p}F[\beta(U-H(\textbf{p}))] \nonumber \\
   &=& \frac{1}{\mathcal{G}} \frac{\partial}{\partial V}
   V^{n/3} \int
   d^n\mathbf{p}F[\beta(U-H(\textbf{p}))]\nonumber \\
   &=&
   \frac{n}{3}\frac{1}{V}
\end{eqnarray}
from which the standard ideal gas law is easily obtained:
\begin{equation}\label{}
    P_\rho V_\rho = \frac{n}{3} T_\rho
\end{equation}
The fact that the standard ideal gas law is found to hold for any
dual statistics should not surprise since it generalizes a result
already found within the nonextensive thermodynamics \cite{Abe01}.
Let us now focus on the form of the function $\beta(U,V)$ in the
ideal gas case. Using the standard change of variable
\cite{Khinchin} $\frac{\mathbf{p}^2}{2m}=e$, $d^n\mathbf{p}= c_n
e^{\frac{n}{2}-1} de$, followed by the change of variable $y=
\beta e$, the condition $<H>=U$ is expressed as:
\begin{equation}\label{eq:<y-A>}
    I(A,n) \doteq \int_0^{\overline{y}(A)} dy y^{\frac{n}{2}-1} (A-y)f(A-y)=0
\end{equation}
where $A= \beta U$, and $\overline{y}$ is a cut-off possibly
infinite and possibly depending on $A$. We shall refer to Eq.
(\ref{eq:<y-A>}) as the energy constraint equation. From such
equation it is easily inferred that, if for a given number of
degrees of freedom $n$, a solution $A_n$ of (\ref{eq:<y-A>})
exists, then the function $\beta(U,V)$ exists and is given by:
\begin{equation}\label{eq:betaIsAnOverE}
    \beta = \frac{A_n}{U}
\end{equation}
For example, within the canonical ensemble one has $\beta =
\frac{n}{2U}$. In the ideal gas case $<K>=U$, so, from the
definition of macroscopic state (\ref{eq:statedef}) one has
\begin{equation}\label{eq:TisNover2U}
T_\rho = \frac{2U}{n}.
\end{equation}
Because of orthodicity we have $\frac{\partial S_\rho}{\partial
U}= \frac{1}{T_\rho} = \frac{n}{2U}$ for any dual ensemble. This
implies that the entropy is:
\begin{equation}\label{eq:sakur}
    S_\rho(U,V) =  \frac{n}{2}\log{U} + \frac{n}{3}\log{V} + const
\end{equation}
Therefore the standard ideal gas thermodynamics is recovered for
any dual orthode. This means that the canonical or microcanonical
statistics are not the only necessary statistics that give the
ideal gas thermodynamics. On the contrary ordinary thermodynamics
may follow from non-ordinary ensembles belonging to the class of
dual orthodes.

As the generalized equipartition theorem (\ref{eq:gen-equi-teo})
suggests, in general the standard relation $T_\rho=
\frac{1}{\beta}$ does not hold. For example, from Eq.s
(\ref{eq:betaIsAnOverE}) and (\ref{eq:TisNover2U}), one easily
finds the following formula:
\begin{equation}\label{}
    T_\rho = \frac{1}{\beta} \frac{\frac{n}{2}}{A_n}
\end{equation}
By comparison with the Eq. (\ref{eq:gen-equi-teo}), one also
deduces that,
\begin{equation}\label{eq:GoverG}
    \frac{\mathcal{G}}{G} = \frac{\frac{n}{2}}{A_n}.
\end{equation}
The previous formula can also be derived directly by considering
the explicit expression of $\mathcal{G}$ (we adopt the $U,V$
representation),
\begin{equation}\label{}
    \mathcal{G}(U,V) = c_n V^{\frac{n}{3}}
    \int_0^{\overline{U}}de e^{\frac{n}{2}-1}F\left[A_n(1+e/U)\right]
\end{equation}
where $\overline{U}$ is the cut-off energy value. Equation
(\ref{eq:GoverG}) follows after an integration by parts and the
definition of $U$ from Eq. (\ref{eq:statedef}).

\section{\label{sec:examples}Examples}
\subsection{\label{subsec:Tsallis}Tsallis ensemble}
As an illustration of the theory let us first apply it to the
Tsallis orthodes \cite{Campisi07} of indices $q$. As we will see,
this is a quite special case that can be worked analytically. The
ensembles are (we adopt the $U,V$ representation):
\begin{equation}\label{}
    \rho_q (\textbf{z};U,V) = \frac{\left[ 1 -
\frac{\beta}{\alpha_q}(H(\textbf{z};V)-U)\right]^{\alpha_q-1}}{N(U,V)}
\end{equation}
where for simplicity we have adopted the notation $\alpha_q =
\frac{1}{1-q}$. In this case the energy constraint integrals
(\ref{eq:<y-A>}) can be evaluated analytically:
\begin{equation}\label{}
    I_q(A,n) = -\left(\frac{n}{2}-A \right)
    \frac{\left(A+\alpha_q\right)^{\alpha_q+n/2-1}}{\alpha_q^{\alpha_q}}\frac{\Gamma(\alpha_q+1)\Gamma(\frac{n}{2})}{\Gamma(\frac{n}{2}+\alpha_q+1)}
\end{equation}
The solutions $A_{n,q}$ of the equations $I_q(A,n)=0$ are $A_{n,q}
= \frac{n}{2}$, no matter the value of $q$. Therefore one finds,
from Eq. (\ref{eq:betaIsAnOverE}), the relation $T^{[q]} =
\frac{1}{\beta}$, which is in agreement with the Tsallis
equipartition theorem (see Eq. (\ref{eq:equi-teo-tsal}))

Using Eq. (\ref{eq:betaIsAnOverE}) one can express the Tsallis
ensemble, in the ideal gas case, as
\begin{equation}\label{}
    \rho_q(\mathbf{z};U,V) = \frac{\left[1+\frac{(1-q)n}{2}(1-\frac{H(\mathbf{z};V)}{U})\right]^{\frac{q}{1-q}}}
    {\int d\mathbf{z}
    \left[1-\frac{(1-q)n}{2}(1-\frac{H(\mathbf{z};V)}{U})\right]^{\frac{q}{1-q}}}\nonumber
\end{equation}
where the $U,V$ parametrization has been adopted. Alternatively,
adopting the $\beta, V$ parametrization the Tsallis ensemble would
read as:
\begin{equation}\label{}
    \rho_q(\mathbf{z};U,V) = \frac{\left[1+ (1-q) (\frac{n}{2}- \beta H(\mathbf{z};V))\right]^{\frac{q}{1-q}}}
    {\int d\mathbf{z}
    \left[1+(1-q) (\frac{n}{2} -\beta H(\mathbf{z};V))\right]^{\frac{q}{1-q}}}\nonumber
\end{equation}
Applying Eq.s (\ref{eq:S=logMathcalG}) and
(\ref{eq:mathcalG=intF}) gives the entropy. In the $U,V$
representation, it reads:
\begin{equation}\label{}
    S^{[q]}(U,V)= \frac{n}{3} \log V +\log L_{n,q}(U) + \log  c_n
\end{equation}
where
\begin{equation}\label{}
   L_{n,q}(U) = U^{\frac{n}{2}}\int_0^{1+\frac{2}{n(1-q)}}dx
   x^{\frac{n}{2}-1}\left[1+\frac{(1-q)n}{2}\left(1-x\right)\right]^{\frac{1}{1-q}}.
\end{equation}
In agreement with Eq. (\ref{eq:sakur}), the dependence of the
entropy on $U$ is of the type $\frac{n}{2}\log U$. The integral
$L_{n,q}$ has been obtained by using the cut-off condition
$\overline{e}= U\left(1+\frac{2}{n(1-q)}\right)$, and the change
of variable $x = e/U$, where $e$ denotes energy.
\subsection{\label{subsec:gauss}Gaussian ensemble}
\begin{figure}
\includegraphics[width=15cm]{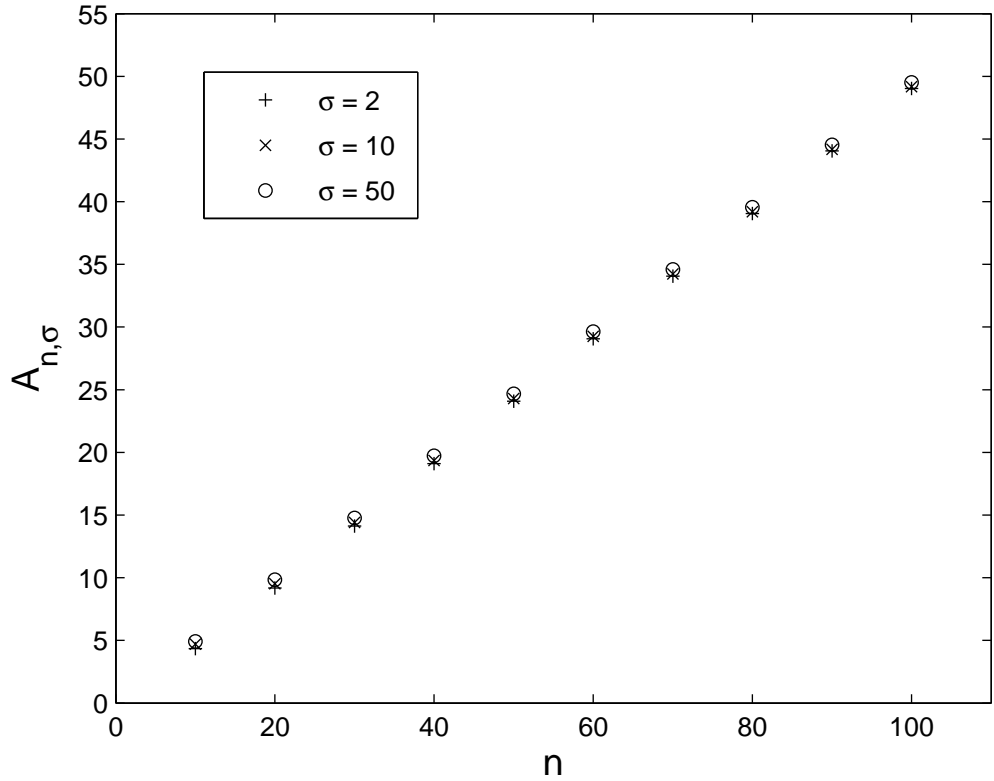}
  \caption{Solution of Eq. (\ref{eq:IsigmaAn}), for different values
  of $\sigma$ and $n$}
\label{fig:An-gaus}
\end{figure}

Since the fundamental work of Khinchin \cite{Khinchin} based on
the application of the central limit theorem, it is known that the
distribution law for a large component of a large Hamiltonian
isolated systems of total energy $a$ is well approximated by a the
following Gaussian distribution:
\begin{equation}\label{eq:Gauss}
    \rho =
    \frac{e^{\beta(a-H)}\exp\left[-\frac{(A_1-H)^{2}}{2B_2}\right]}{normalization},
\end{equation}
the quantities $A_1$ and $B_2$ being defined in terms of the
Laplace transforms $Z_{i}(\beta)$ of the structure functions
$\Omega_{i}(x)$ of the system ($i=1$) and the heat bath ($i=2$):
\begin{eqnarray}\label{}
    A_1 = -\frac{d\log Z_1}{d \beta} \nonumber \\
    B_2 = \frac{d^{2}\log Z_2}{d \beta^{2}} \nonumber \\
\end{eqnarray}
According to Khinchin the quantity $A_1$ is a good approximation
to the average energy $U$ of the system
($\frac{U-A_1}{U}=O(\frac{1}{N_1}))$, where $N_1$ is the number of
particles in the system). Besides $B_2$, the width of the
distribution, can be expressed, in the case of an ideal gas bath
(see Chapter 5, Section 22 of Ref. \cite{Khinchin}) as:
\begin{equation}\label{}
    B_2 = \frac{3N_2}{2 \beta^{2}}
\end{equation}
Here $N_2$ denotes the number of particles in the bath. Hence the
ensemble in Eq. (\ref{eq:Gauss}) can be re-expressed in the form
of a dual ensemble:
\begin{equation}\label{eq:dualGauss}
    \rho_\sigma(\textbf{z};U,V)= e^{\beta(U-H)}\frac{\exp\left[-\frac{(\beta(U-H))^{2}}{2\sigma}\right]}{G}
\end{equation}
where $\sigma = \frac{3N_2}{2}$ is indeed the specific heat of the
heat bath, namely $\sigma$ plays the same role here as the
parameter $\alpha_q= \frac {1}{1-q}$ in the Tsallis ensembles (see
Eq. (\ref{eq:CV-q})). The Gaussian ensemble is reproduced with the
choice:
\begin{eqnarray}\label{}
    f(x) &=& e^{x} \exp\left[-\frac{x^2}{2\sigma}\right] \\
    F(x) &=& \sqrt{\frac{\pi \sigma}{2}}e^{\frac{\sigma}{2}}
    \left(1+\texttt{erf}
    \left[\frac{x-\sigma}{\sqrt{2\sigma}} \right]
    \right)
\end{eqnarray}
The energy constraint condition (\ref{eq:<y-A>}) in this ensemble
is
\begin{equation}\label{eq:IsigmaAn}
    I_\sigma(A)= \int_0^{\infty} dy
    y^{\frac{n}{2}-1}(A-y)e^{A-y-\frac{(A-y)^2}{2\sigma}}=0.
\end{equation}
The solutions $A_{n,\sigma}$ of this equation have been evaluated
numerically for $n=10,20,...,100$, $\sigma = 2,10,20$, and shown
in Figure \ref{fig:An-gaus}.
%------------------------------------------------------------------------------------------------------------------------------figure -----------
%-------------------------------------------------------------------------------------------------------------------------------------------------
For all the values of $\sigma$ investigated, $A_{n,\sigma} \simeq
\frac{n}{2}$. The large $n$ behaviour of $A_{n,\sigma}$ is
investigated in \ref{appAn-gauss}. The entropy, in the $U$
representation is:
\begin{equation}\label{eq:GausEnt}
    S(U,V)=\frac{n}{3}\log V+ \log Y_{n,\sigma}(U) + \log{c_n\sqrt{\frac{\pi \sigma}{2}}e^{\frac{\sigma}{2}}}
\end{equation}
where
\begin{equation}\label{}
    Y_{n,\sigma}(U)=U^{\frac{n}{2}}\int_0^{\infty} dx x^{\frac{n}{2}-1}
    \left(1+\texttt{erf}
    \left[\frac{A_{n,\sigma}(1-x)-\sigma}{\sqrt{2\sigma}} \right]
    \right)
\end{equation}
In agreement with Eq. (\ref{eq:sakur}), the dependence of the
entropy on $U$ is of the type $\frac{n}{2}\log U$. The integral
$Y_{n,\sigma}$ has been obtained by using the change of variable
$x = e/U$, where $e$ denotes energy.

The Gaussian ensemble of Eq. (\ref{eq:dualGauss}) interpolates
between canonical and microcanonical ensembles as does the Tsallis
ensemble \cite{Challa88}. In fact, on one hand $f(x) \rightarrow
e^x$ as $\sigma \rightarrow +\infty$, and, on the other $f(x)
\rightarrow \sqrt{2\pi\sigma}\delta(x)$ as $\sigma \rightarrow 0$
(the vanishing term $\sqrt{2\pi\sigma}$ is not a problem because
it will be cancelled with the normalization).

\begin{figure}
\includegraphics[width=15cm]{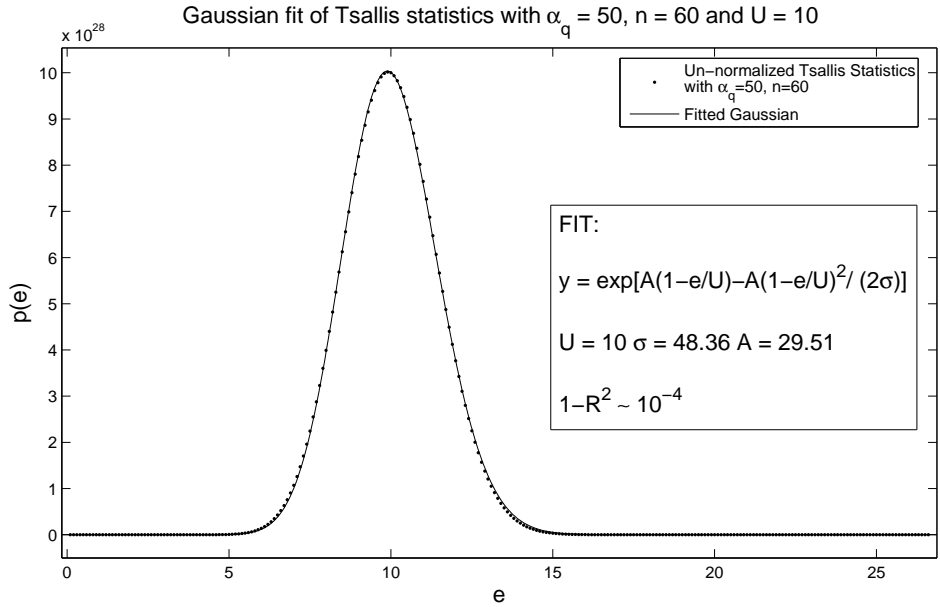}
  \caption{Gaussian fit of Tsallis statistics for $\alpha_q = 50, n= 60, U = 10$. The dual Gaussian ensemble approximates Tsallis ensembles for large values of $n$ and $\alpha_q$}
\label{fig:FitTsalGaus}
\end{figure}

Further, based on the fact that the Gaussian ensemble of index
$\sigma$ well describes the statistics of a large component of a
large isolated system, we deduce that it must well approximate the
Tsallis statistics of index $\alpha_q = \sigma$ in the case
$n,\alpha_q \gg 1$. This fact is illustrated in Figure
\ref{fig:FitTsalGaus}, where we have plotted the unnormalized
Tsallis distribution
$\left(1+\frac{n}{2\alpha_q}(1-\frac{e}{U})\right)^{\alpha_q-1}e^{\frac{n}{2}-1}$,
for $\alpha_q = 50, n= 60, U = 10$, and fitted it to the
unnormalized gaussian statistics
$\exp[A(1-e/U)-\frac{(A(1-e/U))^2}{2\sigma}]e^{\frac{n}{2}-1}$
with $U,\sigma, A$ as free parameters. The fit is very good
($1-R^2 = O(10^{-4})$) and the fitting parameter matched quite
well the expected values: $\sigma = 48.36 \simeq \alpha_q=50$, $U
= 10$, $A = 29.51 \simeq A_{n=60,\sigma=50} = 29.63$.
%------------------------------------------------------------------------------------------------------------------------------figure -----------
%-------------------------------------------------------------------------------------------------------------------------------------------------

\subsection{\label{subsec:Fermi}Fermi-like ensemble}
\begin{figure}
\includegraphics[width=15cm]{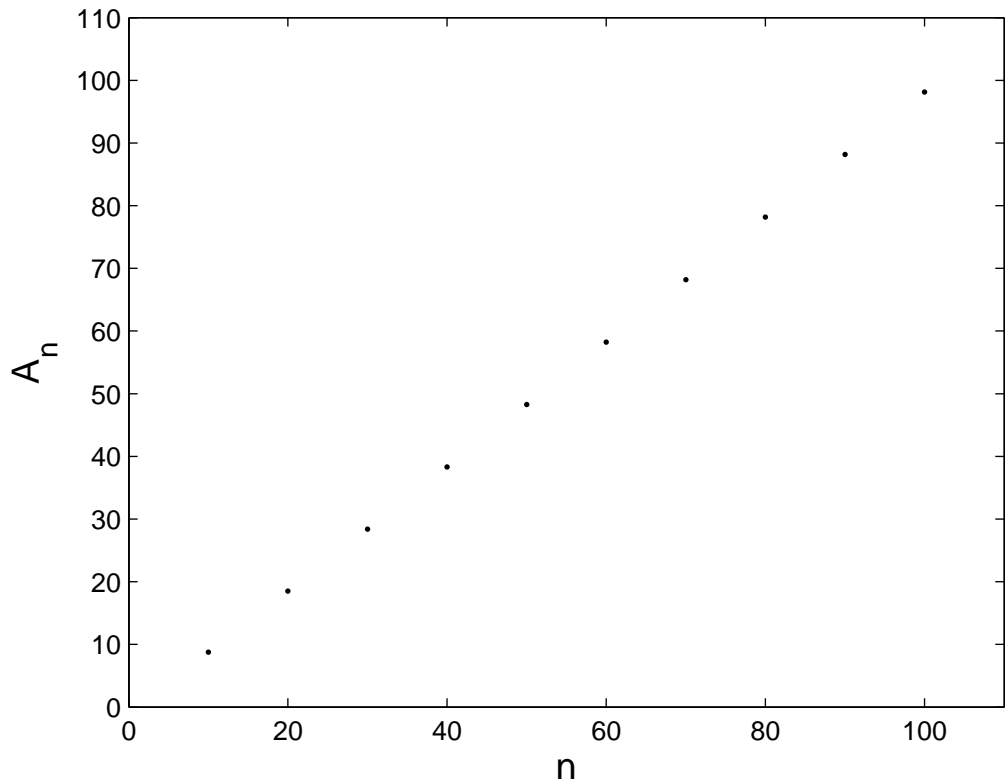}
  \caption{Solution of Eq. (\ref{eq:IAffermi}) for different values
  of $n$.}
\label{fig:Af-fermi}
\end{figure}

The theory developed so far allows us to construct mechanical
models of thermodynamics with the most diverse types of
distributions. For example one may ask whether it would be
possible to have an ensemble with a Fermi-like distribution. This
is possible for the ideal gas case. We will describe this ensemble
as an illustration of the theory, without discussing whether it
really applies to some many-particle physical system. The
Fermi-like statistics uses the choice:
\begin{eqnarray}\label{}
    F(x) &=& \log (e^{x}+1) \\
    f(x) &=& \frac{1}{e^{-x}+1}
\end{eqnarray}
with no finite cut-off. The solutions $A_n$ of the energy
constraint equation (\ref{eq:<y-A>}):
\begin{equation}\label{eq:IAffermi}
    I (A,n) = \int_0^{\infty} dy  \frac{y^{\frac{n}{2}-1}
    (A-y)}{e^{y-A}+1}=0
\end{equation}
have been evaluated numerically and shown in Fig.
\ref{fig:Af-fermi} for the values $n=10,20,...,100$. From the
figure we see that, $A_n \simeq n$. This fact is analyzed in more
details in \ref{appSec:An-fermi}
%------------------------------------------------------------------------------------------------------------------------------figure -----------
%-------------------------------------------------------------------------------------------------------------------------------------------------

Adopting the $(U,V)$ parametrization, the ensemble, is:
\begin{equation}\label{}
    \rho(\mathbf{z};U,V) = \frac{ \left[ e^{-A_n(1-\frac{H(\mathbf{z};V)}{U})}+1 \right]^{-1}}
            {\int d \mathbf{z} \left[ e^{-A_n(1-\frac{H(\mathbf{z};V)}{U})}+1 \right]^{-1}}
\end{equation}
or, alternatively:
\begin{equation}\label{}
    \rho(\mathbf{z};\beta,V) = \frac{ \left[ e^{-A_n+ \beta H(\mathbf{z};V)}+1 \right]^{-1}}
            {\int d \mathbf{z} \left[ e^{-A_n+ \beta H(\mathbf{z};V)}+1
            \right]^{-1}}
\end{equation}
if the $(\beta,V)$ parametrization is adopted. Using Eq.
(\ref{eq:S=logMathcalG}) gives the entropy. Adopting the $(U,V)$
parametrization, this is given by
\begin{equation}\label{eq:FermiEnt}
    S(U,V) = \log{c_n}+ \frac{n}{3}\log{V} + \log{J_n(U)}
\end{equation}
where
\begin{equation}\label{}
    J_n(U) = U^{\frac{n}{2}}\int_0^{\infty} dx x^{\frac{n}{2}-1}
    \log\left(e^{A_n(1-x)}+1\right)
\end{equation}

In agreement with Eq. (\ref{eq:sakur}), the dependence of the
entropy on $U$ is of the type $\frac{n}{2}\log U$. The integral
$J_{n}$ has been obtained by using the change of variable $x =
e/U$, where $e$ denotes energy.

\section{\label{sec:conclusions}Discussion and conclusion}
\begin{figure}
\includegraphics[width=15cm]{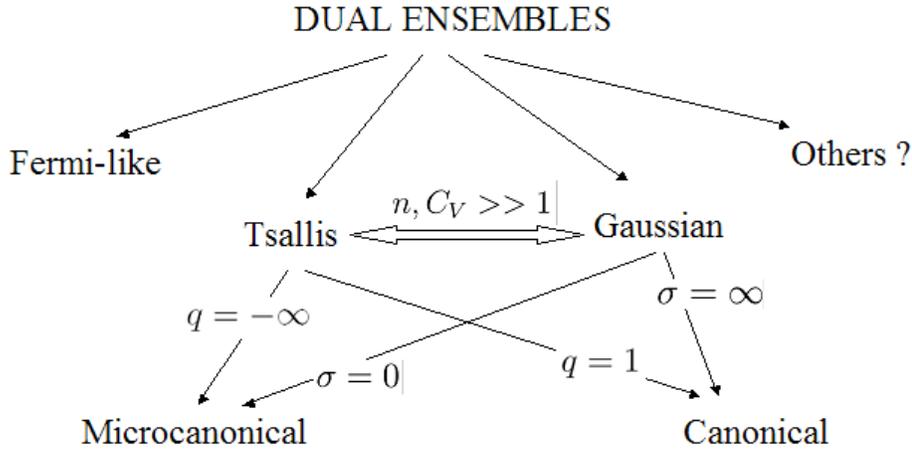}
  \caption{Generalization scheme. Tsallis ensembles, Gaussian ensembles, Fermi-like ensemble, and possibly
  other kinds of ensembles belong to the class of dual orthodes. Canonical and microcanonical are special instances of Tsallis and Gaussian ensembles.
  For large systems ($n>>1$) Gaussian and Tsallis ensemble coincide if $\sigma=\frac{1}{1-q} = C_V >>
  1$.}
\label{fig:schema}
\end{figure}

In this work we addressed some fundamental issues raised recently
in statistical mechanics, namely whether a theoretical basis can
be provided for non-standard (i.e. neither canonical nor
microcanonical) ensembles, which are often encountered in the most
diverse fields of physics. Following a line initiated in Refs.
\cite{Campisi07,Campisi06.2},  we used Boltzmann's original
approach based on the ``heat theorem'', in order to provide a
fresh look on the subject. By generalizing the \emph{duality}
property observed in the Tsallis case, we have been able to define
the class of \emph{dual statistics}, which includes the Tsallis
ensembles as particular cases. The generalization scheme is
represented in Fig. \ref{fig:schema}.
%------------------------------------------------------------------------------------------------------------------------------figure -----------
%-------------------------------------------------------------------------------------------------------------------------
Thanks to the proposed generalization it is possible to provide a
theoretical basis for many non-standard statistics other than
Tsallis', such as the gaussian and the Fermi-like statistics. For
all such non-standard orthodes the heat theorem holds, namely the
usual thermodynamic relations are recovered. In this paper we also
have provided a general formula for the entropy associated to any
dual orthode. This allowed us to write down the expression of
entropy for the Gaussian ensemble, which has never been done
before. To use the same expression as Gallavotti
\cite{Gallavotti}, all dual orthodes provide ``mechanical models
of thermodynamics''. This result is not trivial since the only
known orthodes were the microcanonical ensemble, the canonical
ensemble and its variants like the grand canonical ensemble and
the pressure ensemble \cite{Gallavotti}. Despite the fact that the
class of orthodes is quite vast, the microcanonical and the
canonical ensembles still play a special role in statistical
mechanics because they are cases of ``hidden dual statistics''.
They do not rely on the employment of the energy constraint, which
constitutes the mechanism through which it is possible to
construct non-standard dual orthodes.

One interesting result is that the classical ideal gas
thermodynamics follows from any dual ensemble, thus revealing that
this occurrence is not special for the standard statistics. This
fact was already noticed for the Tsallis case a few years ago
\cite{Abe01}, although its orthodicity was not yet clearly
recognized. It also suggests that there is a one-to-one
correspondence of states obtained in different dual ensembles,
i.e. there is an \emph{equivalence} of all dual ensembles. Such
equivalence holds no matter the number of degrees of freedom, and,
of course, may break as one considers more realistic Hamiltonian
models of systems with phase transitions. We also like to stress
that all dual orthodes are equivalent also in another sense,
namely the sense first investigated by Gibbs \cite{Gibbs02} for
the canonical and microcanonical ensembles. In the thermodynamic
limit ($n \rightarrow \infty$) and for the free gas model, the
canonical ensemble is so peaked around the average energy value
that it is practically undistinguishable from the microcanonical
one \cite{Huang}. The same kind of equivalence occurs for any dual
statistics provided that $A_n \rightarrow \infty$ in the
thermodynamic limit, which is indeed the case observed in the
examples considered in this paper. This is because the
distribution is expressed in terms of the quantity
$f\left(A_n\frac{U-e}{U}\right)$ which tends to the Dirac delta
(times an unimportant proportionality factor), centered around
$U$:
\begin{equation}\label{}
    f\left(A_n\frac{U-H}{U}\right) \rightarrow const \times
    \delta(U-H)
\end{equation}
This follows from the asymptotic formula $h(kx) \rightarrow
\frac{\int dy h(y)}{k} \delta(x)$, as $k \rightarrow \infty$.

\section*{Acknowledgements}
The author wishes to thank Prof. D. H. Kobe for the many useful
comments on the manuscript and his constant encouragement.

\appendix
\section{\label{appSec:gen-equi-theo}Proof of the Generalized Equipartition Theorem of Eq. (\ref{eq:gen-equi-teo})}
Let us calculate explicitly the average value of
$p_i\frac{\partial H}{\partial p_i}$. Regardless of the
parametrization one has:
\begin{eqnarray}\label{}
    \left\langle p_i \frac{\partial H}{\partial p_i}
 \right\rangle_\rho &=& \frac{1}{G} \int d\textbf{z} p_i \frac{\partial H}{\partial
 p_i}f[\beta(U-H)] \nonumber \\
 &=& -\frac{1}{\beta G} \int d\textbf{z} p_i \frac{\partial}{\partial
 p_i}F[\beta(U-H)] \nonumber \\
 &=& -\frac{1}{\beta G}
 \left[ \left[p_i F[\beta(U-H(\textbf{z};V))]\right]_{\textbf{z} \in \partial \mathcal{D}}
- \int d\textbf{z} F[\beta(U-H)]
 \right] \nonumber \\
\end{eqnarray}
Where we integrated by parts over $p_i$ to obtain the third line.
From Eq.s (\ref{eq:mathcalG=intF}) and (\ref{eq:F-cutoff})
follows:
\begin{equation}\label{}
    \left\langle p_i \frac{\partial H}{\partial p_i}
 \right\rangle_\rho = \frac{1}{\beta}\frac{\mathcal{G}}{G}
\end{equation}
According to the parametrization adopted this would be either Eq.
(\ref{eq:genEquiTheoUV}) or (\ref{eq:genEquiTheoBETAV}).

\section{\label{appSec:orthodicity}Proof of orthodicity in the $\beta,V$ parametrization}
In the $\beta,V$ parametrization $U$ is a function of $(\beta,V)$.
Therefore $F$ is a function of $(\mathbf{z};U,V)$, and $\mathcal
G$ (Eq. \ref{eq:mathcalG=intF}) and $S_{\rho}$ (Eq.
\ref{eq:S=logMathcalG}) are functions of $(\beta,V)$. Let us
calculate the partial derivatives of $S$
 \begin{eqnarray}\label{}
    \frac{\partial S_\rho}{\partial \beta}  &=& \frac{1}{\mathcal{G}}
    \frac{\partial}{ \partial \beta}\int
d\textbf{z} F(\beta(U-H))\nonumber \\
    &=& \frac{1}{\mathcal{G}}\int d\textbf{z} f(\beta(U-H)) \left[ (U-H) + \beta\frac{\partial
U}{\partial \beta}
    \right]\nonumber \\
    &=& - \frac{G}{\mathcal{G}} \langle H-U \rangle_\rho +
    \frac{G}{\mathcal{G}} \beta\frac{\partial U}{\partial \beta} = \frac{1}{T_\rho}\frac{\partial U_{\rho}}{\partial \beta}
\end{eqnarray}
where we used the equation $U_\rho = <H>_{\rho}$ from the state
definition (\ref{eq:statedef}) and the generalized equipartition
theorem of Eq. (\ref{eq:gen-equi-teo}).
 \begin{eqnarray}\label{}
    \frac{\partial S_\rho}{\partial V_\rho} &=& \frac{1}{\mathcal{G}}
    \frac{\partial}{\partial V}\int
d\textbf{z} F(\beta(U-H))\nonumber \\
    &=& \frac{1}{\mathcal{G}}\int d\textbf{z} f(\beta(U-H)) \left[
    \beta \frac{\partial U}{\partial V} - \beta \frac{\partial H}{\partial V} \nonumber \right]\\
    &=& \beta \frac{G}{\mathcal{G}} \frac{\partial U}{\partial V}
    -
    \frac{G}{\mathcal{G}}\beta   \left\langle \frac{\partial H}{\partial V} \right\rangle_\rho
\nonumber \\
     &=& \frac{1}{T_{\rho}}\frac{\partial U_{\rho}}{\partial V_{\rho}}    +   \frac{P_\rho}{T_\rho}
\end{eqnarray}
where we used the expression for $P_{\rho}$ from Eq.
(\ref{eq:statedef}) and the generalized equipartition theorem of
Eq. (\ref{eq:gen-equi-teo}). Combining all together:
\begin{equation}\label{}
    dS = \frac{\frac{\partial U_{\rho}}{\partial \beta}d\beta + \frac{\partial U_{\rho}}{\partial V_{\rho}}dV_{\rho} + P_\rho
    dV_{\rho}}{T_{\rho}} = \frac{dU_{\rho} + P_\rho
    dV_{\rho}}{T_{\rho}}
\end{equation}

\section{\label{appAn-gauss}Large $n$ behaviour of the coefficients $ A_{n,\sigma} $}

Eq. (\ref{eq:IsigmaAn}) can be recast in the following form:
\begin{equation}\label{eq:C1}
    \int_0^{\infty} dy g_n(y) e^{-\frac{(A-y)^2}{2 \sigma}} = A \int_0^{\infty} dy g_n(y) y^{-1} e^{-\frac{(A-y)^2}{2 \sigma}}
\end{equation}
where
\begin{equation}\label{}
    g_n(y) \doteq \frac{y^{\frac{n}{2}} e^{-y} }{\int_0^{\infty} y^{\frac{n}{2}}
    e^{-y}}\theta(y)
\end{equation}
and $\theta$ is Heaviside's step function. Its Fourier transform
is:
\begin{equation}\label{}
    \hat{g}_n(k) = (1-ik)^{-\frac{n}{2}-1}
\end{equation}
Using the formula $\lim_{N \rightarrow \infty}\left(1+\frac{x}{N}
\right)^{N} = e^x$, we have:
\begin{equation}\label{}
    \lim_{n \rightarrow \infty} \hat{g}_n(k) = e^{ik\left(\frac{n}{2}+1\right)}
\end{equation}
therefore:
\begin{equation}\label{}
    \lim_{n \rightarrow \infty} g_n(y) = \delta
    \left(y-\left(\frac{n}{2}+1\right) \right).
\end{equation}
Using this result Eq. (\ref{eq:C1}), becomes, for very large $n$:
\begin{equation}\label{}
    A \simeq \frac{n}{2}
\end{equation}
This result does not depend on $\sigma$.

\section{\label{appSec:An-fermi}Large $n$ behaviour of the coefficients $A_n$ for the Fermi-like case}
Figure \ref{fig:Af-fermi} suggests that, in the limit $n
\rightarrow \infty$, $A_n \simeq n$. In this appendix section we
provide a simple consistency argument to support the claim that
$A_n \simeq n$. Let us assume that for very large $n$, $A_n \simeq
n$. Then we should have from Eq. (\ref{eq:IAffermi}):
\begin{equation}\label{eq:approx-solution-fermi}
     \int_0^{\infty} dy  \frac{y^{\frac{n}{2}}}{e^{y-n}+1} \simeq
     n \int_0^{\infty} dy  \frac{y^{\frac{n}{2}-1}} {e^{y-n}+1}
\end{equation}
Equating the first derivative of the integrand in the left hand
side to zero, gives:
\begin{equation}\label{}
    y = \frac{n}{2}(1+e^{n-y})  \nonumber
\end{equation}
which is satisfied for $y=n$. This means that an extremum (a
maximum as we will see) is attained for $y=n$. The value taken by
the integrand at the maximum is $\frac{n^{\frac{n}{2}}}{2}$ which
increases very quickly. The second derivative, calculated at
$y=n$, is:
\begin{equation}\label{}
    \frac{1}{4}\left[ \left(\frac{n}{2}-1 \right)n^{-\frac{n}{2}+1}- n^{\frac{n}{2}} \right]
\end{equation}
which tends to $-\infty$ very quickly. This indicates that the
integrand becomes very sharply peaked around $y=n$ as $n$
increases. Therefore, as an approximation, we can replace $y$ with
$n$, and see that $y^{\frac{n}{2}} \simeq ny^{\frac{n}{2}-1}$,
thus getting Eq. (\ref{eq:approx-solution-fermi}).

\section*{References}
\bibliographystyle{unsrt}
\bibliography{thebibliography}% Produces the bibliography via BibTeX.

\end{document}